\documentclass[conference]{IEEEtran}
\IEEEoverridecommandlockouts
\usepackage{amsmath,amssymb,amsfonts}
\usepackage{algorithmic}
\usepackage{graphicx}
\usepackage{booktabs}
\usepackage{multirow}
\usepackage{subfig}
\usepackage{float}
\usepackage{cite}
\usepackage{makecell}
\usepackage{textcomp}
\usepackage{xcolor}
\usepackage[ruled,linesnumbered]{algorithm2e}
\DeclareRobustCommand*{\IEEEauthorrefmark}[1]{%
    \raisebox{0pt}[0pt][0pt]{\textsuperscript{\footnotesize\ensuremath{#1}}}}
\def\BibTeX{{\rm B\kern-.05em{\sc i\kern-.025em b}\kern-.08em
    T\kern-.1667em\lower.7ex\hbox{E}\kern-.125emX}}
\begin{document}

\title{A Wi-Fi Signal-Based Human Activity Recognition Using High-Dimensional Factor Models}

\author{
\IEEEauthorblockN{
Junshuo Liu\IEEEauthorrefmark{1}, Fuhai Wang\IEEEauthorrefmark{1}, Zhe Li\IEEEauthorrefmark{1}, Rujing Xiong\IEEEauthorrefmark{1}, Tiebin Mi\IEEEauthorrefmark{1}, 
Robert Caiming Qiu\IEEEauthorrefmark{1}}
\IEEEauthorblockA{\IEEEauthorrefmark{1}School of Electronic Information and Communications, Huazhong University of Science and Technology \\ Wuhan 430074, China}
Emails: \{junshuo\_liu, wangfuhai, m202272435, rujing, mitiebin, caiming\}@hust.edu.cn
}

\maketitle

\begin{abstract}
Passive sensing techniques based on Wi-Fi signals have emerged as a promising technology in advanced wireless communication systems due to their widespread application and cost-effectiveness. However, the proliferation of low-cost Internet of Things (IoT) devices has led to dense network deployments, resulting in increased levels of noise and interference in Wi-Fi environments. This, in turn, leads to noisy and redundant Channel State Information (CSI) data. As a consequence, the accuracy of human activity recognition based on Wi-Fi signals is compromised. To address this issue, we propose a novel CSI data signal extraction method. We established a human activity recognition system based on the Intel 5300 network interface cards (NICs) and collected a dataset containing six categories of human activities. Using our approach, signals extracted from the CSI data serve as inputs to machine learning (ML) classification algorithms to evaluate classification performance. In comparison to ML methods based on Principal Component Analysis (PCA), our proposed High-Dimensional Factor Model (HDFM) method improves recognition accuracy by 6.8\%.
\end{abstract}


\section{Introduction}
Human Activity Recognition (HAR) plays a pivotal role in the Internet of Things (IoT), particularly in smart homes and offices\cite{jovanov2019wearables}. HAR involves categorizing human activities, offering applications in elderly care, healthcare management, and home security\cite{hassan2018robust,zhou2020deep,zhao2022human,lara2012survey}.

HAR techniques are categorized into three domains based on sensing modalities: 1) Vision-based HAR\cite{xu2013exploring,kim2019vision,sharma2022review}; 2) Wearable device-based HAR\cite{bhat2018online,qiu2022multi,zhang2022deep}; and 3) Wireless sensing-based HAR\cite{zhu2022continuous,oguntala2019smartwall,wang2017device,yan2019wiact}. Vision-based HAR relies on high-resolution cameras but faces challenges in non-line-of-sight scenarios and privacy concerns\cite{sun2022human}. Wearable device-based HAR uses integrated sensors, including accelerometers, magnetometers, gyroscopes, barometers, and physiological sensors\cite{qiu2022multi}, for data collection and activity classification, but may impose a burden on users, especially the elderly and children\cite{sun2022human}.

The wireless sensing-based HAR domain encompasses various technologies, including radar, radio frequency identification (RFID), and wireless local area network (WLAN). Radar sensors capture human motion without the need for dedicated devices and are robust under varying lighting conditions. Their extensive operational bandwidth provides exceptional range resolution capabilities\cite{islam2022human}. However, radar-based HAR systems face limitations such as reduced portability and higher hardware costs. RFID-based systems offer a cost-effective solution, requiring only a chip tag and an antenna. Nonetheless, they encounter challenges in ensuring reliability in complex indoor environments, mainly due to issues like multipath fading and temporal dynamics\cite{yang2013rssi}. Conversely, WLAN technology enables user connectivity via Wi-Fi signals, presenting economic and energy-efficient advantages when compared to radar. It is unaffected by lighting conditions, simpler to implement, and raises fewer privacy concerns than camera-based systems. Additionally, Wi-Fi signals exhibit resilience in non-line-of-sight conditions.

Similar to RFID technology, Wi-Fi devices utilize Received Signal Strength Indication (RSSI). However, RSSI's instability makes it suboptimal for accurately capturing dynamic signal variations\cite{yang2013rssi}. In contrast, Wi-Fi systems can access Channel State Information (CSI), which has gained attention for its abundant and stable information content\cite{halperin2011tool}. Unlike RSSI, CSI provides both amplitude and phase information across multiple channels and orthogonal frequency division multiplexing (OFDM) subcarriers. During specific activities near the transmitter and receiver, reflected wireless signals from the human body create distinctive patterns. However, raw CSI measurements in real-world Wi-Fi systems often suffer from amplitude noise and phase offsets due to hardware and software errors\cite{chen2023cross,ma2019wifi}. Phase offset removal, such as unwrapping CSI phases, is particularly valuable for binary and multi-class classification tasks. A simple solution involves mitigating random phase offsets through a linear transformation\cite{wang2015phasefi}. Environmental noise can impede the effective representation of diverse human activities by raw CSI measurements. Consequently, preprocessing is required to remove noise through feature extraction\cite{wang2017device}.

Traditional machine learning (ML) methods have found practical applications in the field of HAR. ML presents a powerful modeling approach, where predictions from various base models, such as decision trees, are aggregated to enhance prediction accuracy. Given the sensitivity of CSI data to changes in indoor signal propagation conditions, with signal characteristics evolving over time, ML offers robust and reliable results.

This paper introduces a new signal extraction method and verifies the effectiveness of the proposed approach for our dataset. The dataset is collected by Intel 5300 NICs and comprises various activities, such as lying down, picking up, sitting down, standing up, and walking.

Overall, the main contributions of this paper can be summarized as follows:
\begin{itemize}
\item[$\bullet$] We define a Wi-Fi passive sensing problem using CSI data and perform preprocessing on the dataset. Effective information is extracted from the initial CSI data through preprocessing, resulting in a feature array that includes both CSI amplitude and phase. Subsequently, action labels are assigned to this dataset.
\item[$\bullet$] To improve activity recognition, we introduce a new signal extraction method based on the Marčenko-Pastur law. This approach shows higher performance advantages under the final experimental training.
\end{itemize}

\section{Principle of Wi-Fi sensing}\label{section2}
\subsection{Channel state information}
Channel state information is a vital metric for characterizing communication links, offering detailed insights into multipath propagation dynamics, including amplitude and phase variations across subcarriers in the frequency domain. It considers factors such as temporal delays, amplitude attenuation, and phase shifts, making it valuable for activity recognition due to its sensitivity to environmental conditions.

In wireless communication, the channel impulse response (CIR) is used to characterize multipath effects under the assumption of linear time invariance. The CIR is mathematically expressed as:
\begin{equation}
\label{CIR}
h(\tau)=\sum_{n=1}^{N}a_{n}e^{-j\theta_{n}}\delta(\tau-\tau_{n})
\end{equation}
where $h(\tau)$ represents the channel impulse response, $N$ is the number of multipath components, $a_{n}$ denotes the complex amplitude of the $n$-th multipath component, $\theta_{n}$ signifies the phase shift for the $n$-th component, $\tau_{n}$ indicates the associated time delay, and $\delta(\tau)$ accounts for temporal alignment with the Dirac delta function.

\begin{figure}[tbp]
  \centering
  \includegraphics[width=0.85\columnwidth]{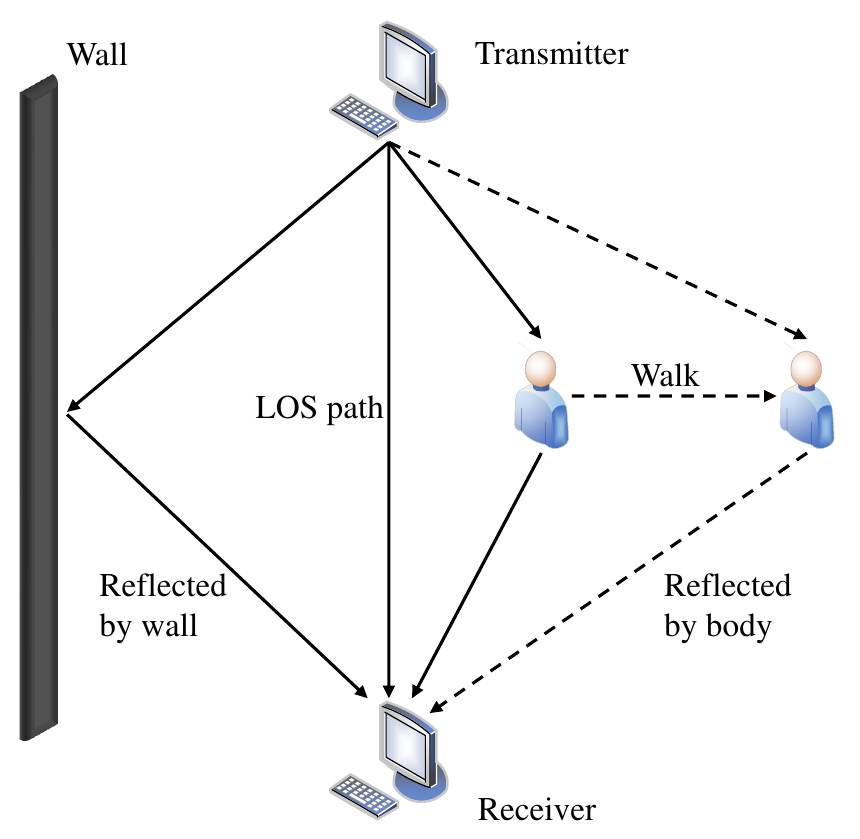}
  \caption{A common scenario illustrating signal propagation within indoor environments.}
  \label{F1}
\end{figure}

When obstructions affect the wireless communication path between a transmitter and receiver, various multipath effects, including reflection, scattering, and fading, impact the received signal, as depicted in Fig.~\ref{F1}. In the IEEE 802.11n OFDM system, the received signal can be expressed as\cite{ma2019wifi}:
\begin{equation}
\label{wirelesschannel}
\mathbf{Y}=\mathbf{H}\mathbf{X}+\mathbf{N}
\end{equation}
where $\mathbf{Y}$ is the received signal vector, $\mathbf{X}$ denotes the transmitted signal vector, $\mathbf{N}$ represents a noise vector, and $\mathbf{H}$ stands for the channel frequency response, which is expressed as a matrix of complex numbers.

CSI characterizes signal attenuation on each transmission path through the channel gain matrix $\mathbf{H}$, encompassing signal scattering, multipath fading, and distance-dependent power decay. Multipath propagation introduces time delay spread in the time domain and selective signal fading in the frequency domain. This multipath effect is described in wireless channels using the channel frequency response (CFR). Assuming unlimited bandwidth, CFR and CIR are linked through the Fourier transform, and the CFR can be represented as:
\begin{equation}
\label{CFR}
H(i)=\Vert H(i) \Vert e^{j\angle H(i)}
\end{equation}
where $H(i)$ corresponds to the CSI of $i$-th subcarrier, $\Vert H(i) \Vert$ denotes the amplitude of the $i$-th subcarrier, and $\angle H(i)$ is the phase shift of the $i$-th subcarrier.

Wi-Fi network interface cards capture CSI measurements, where each measurement comprises $N_{\mathrm{sc}}$ matrices of dimensions $N_{\mathrm{Tx}}\times N_{\mathrm{Rx}}$. Here, $N_{\mathrm{sc}}$, $N_{\mathrm{Tx}}$, and $N_{\mathrm{Rx}}$ represent the counts of OFDM subcarriers, transmitting antennas, and receiving antennas, respectively. Each element in the CSI matrix represents a CFR value characterizing the interaction between transmitting and receiving antennas at a specific OFDM subcarrier in a received Wi-Fi frame. The Intel Wi-Fi Link 5300 NIC, despite operating within an 802.11n OFDM system with 56 subcarriers, exports 30 subcarriers for each of its three antennas via the device driver \cite{halperin2011tool}. Consequently, the NIC reports CSI for 3$\times$3 transmitter-receiver links, each involving 30 subcarriers, resulting in a total of 270 complex CSI values per frame.

\subsection{Preprocessing and Feature Extraction of CSI}
Low-cost commercial Wi-Fi cards introduce various types of noise into CSI measurements, represented as complex matrices subsequently converted into amplitude and phase information. The CSI amplitude undergoes substantial fluctuations due to power adaptations and environmental noise, leading to high-energy impulses and bursts that mask subtle changes induced by human movements. Moreover, the absence of synchronization between the receiver and transmitter's sampling clocks and frequencies introduces Sampling Time/Frequency Offsets (STO/SFO), causing random phase variations within the CSI data\cite{ma2019wifi}. Consequently, extracting meaningful insights from noisy CSI measurements poses a considerable challenge.

Signal extraction eliminates redundant signals from raw CSI measurements, preserving efficient signals through dimension reduction techniques like principal component analysis (PCA) and signal decomposition algorithms\cite{lu2019towards}. Since the time series of CSI subcarriers are highly correlated, PCA is commonly employed to extract linearly uncorrelated variables, known as principal components, through orthogonal transformation\cite{zhang2019widigr,tan2020enabling}. Signal extraction enhances feature extraction for classification algorithms by removing redundant components and capturing information more sensitive to human activity recognition.

In summary, as depicted in Fig.~\ref{F3}, the acquired CSI data necessitates preprocessing to extract pertinent features. These extracted features significantly enhance the effectiveness of algorithmic training, ultimately facilitating the achievement of Wi-Fi sensing objectives. The preprocessing phase predominantly involves two key components: the denoising and conversion of the CSI data and the subsequent signal extraction process.

\begin{figure}[tbp]
  \centering
  \includegraphics[width=1.0\columnwidth]{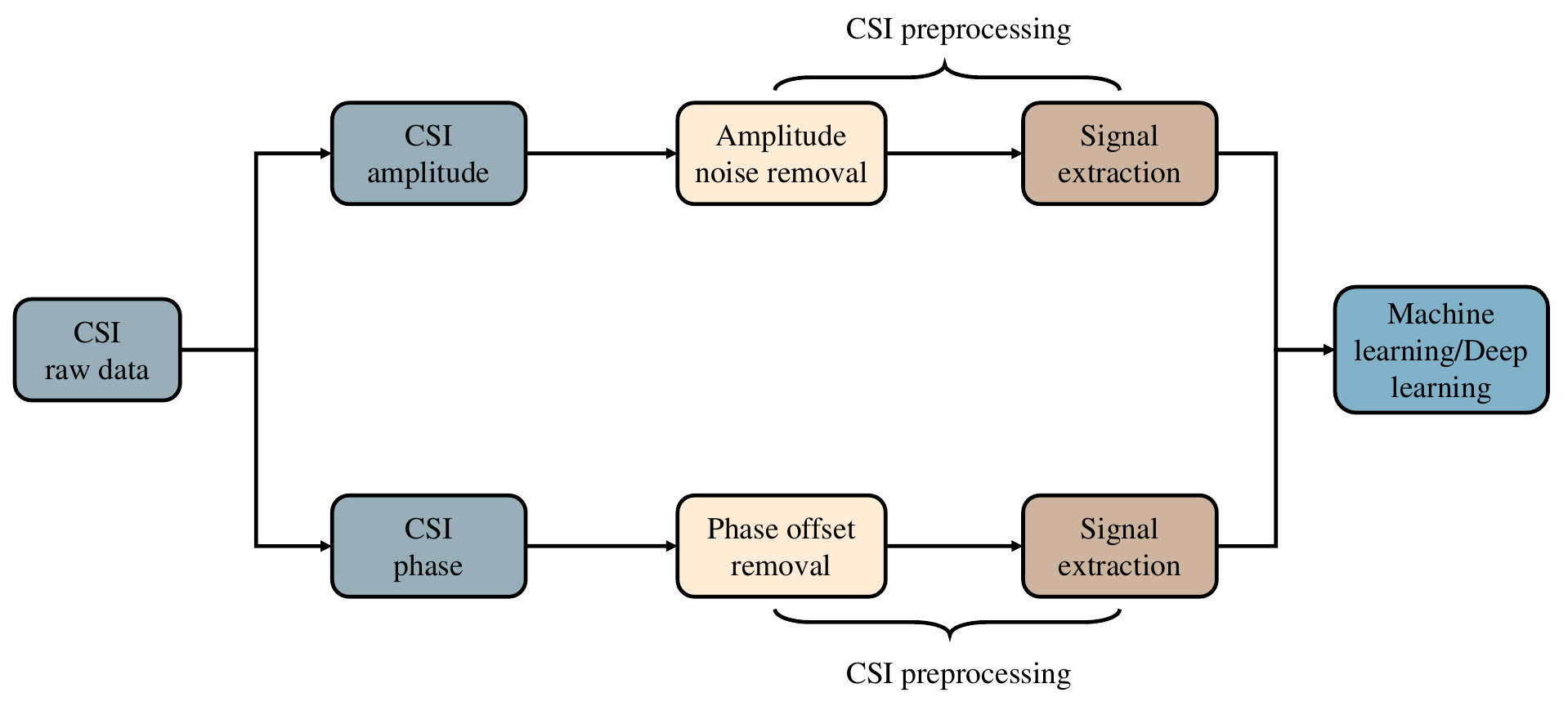}
  \caption{Preprocessing of CSI Data}
  \label{F2}
\end{figure}

\textbf{Amplitude preprocessing.} The internal state changes in Wi-Fi devices often introduce impulse and burst noises in CSI amplitude values. To mitigate high-frequency noise, conventional methods involve the use of Low-pass filters\cite{ma2019practical}, Band-pass filters\cite{zhang2019widigr}, or Butterworth filters\cite{sheng2020deep}. While effective at higher frequencies, these approaches struggle with impulse and bursty noises. Consequently, we propose a high-dimensional factor model (HDFM) for extracting CSI amplitude signals. This method will be explained in Sec.~\ref{Section3} in detail.

\textbf{Phase preprocessing.} Phase information of the $i$-th subcarrier can be extracted from the raw CSI data and is expressed as follows:
\begin{equation}
\hat{\theta}_i = \theta_i + 2\pi \frac{m_i}{N_F} \Delta t + \beta + Z
\label{rawCSI}
\end{equation}
where $\theta_i$ represents the genuine phase, $\Delta t$ signifies the time lag, $\beta$ accounts for a random phase offset induced by unknown factors, $Z$ denotes random measurement noise, and $m_i$ is the subcarrier index for the $i$-th subcarrier. Additionally, $N_F$ corresponds to the FFT (Fast Fourier Transform) size stipulated in the IEEE 802.11n specification \cite{perahia2013next}.

While it is impossible to directly obtain the genuine phase information due to the presence of unknown variables $\Delta t$ and $\beta$, a linear transformation can be applied to the raw phases across the entire frequency band to mitigate the impact of these terms \cite{qian2014pads}.

Let $k$ represent the slope of the phase and $b$ denote the offset across the entire frequency band. Given the 30 subcarriers provided by the Intel 5300 NIC, we have
\begin{equation}
k = \frac{\hat{\theta}_{30}-\hat{\theta}_1}{m_{30}-m_1}, b = \frac{1}{30} \sum_{i=1}^{30} \hat{\theta}_i
\end{equation}

By subtracting $km_i + b$ from the raw phase $\hat{\theta}_i$, we can derive the calibrated phase $\widetilde{\theta}_i$, which can be expressed as:
\begin{equation}
\widetilde{\theta}_i = \hat{\theta}_i - km_i - b
\end{equation}

Since the indices of the 30 subcarriers are symmetric (ranging from -28 to 28 as specified in IEEE 802.11n), it follows that $\sum_{i=1}^{30}m_i=0$. As per Equation (\ref{rawCSI}), the calculation of the offset across the entire frequency band is defined as $b=\frac{1}{30}\sum_{i=1}^{30}\theta_i + \beta + Z$. With the incorporation of the phase slope, offset, and the measured phase of subcarrier $i$, the calibrated phase can be expressed as:
\begin{equation}
\widetilde{\theta}_i = \theta_i - \frac{\theta_{30}-\theta_1}{m_{30}-m_1}m_i - \frac{1}{30}\sum_{i=1}^{30}\theta_i
\end{equation}

This method effectively mitigates phase shifts induced by STO/SFO, enabling the attainment of more accurate phase values. Subsequently, the proposed high-dimensional factor model (HDFM) is employed to eliminate extraneous or redundant components from the received signal.

\section{Proposed methodology and system}\label{Section3}
\subsection{System overview}
\begin{figure*}[tbp]
    \centering
    \includegraphics[width=1.6\columnwidth]{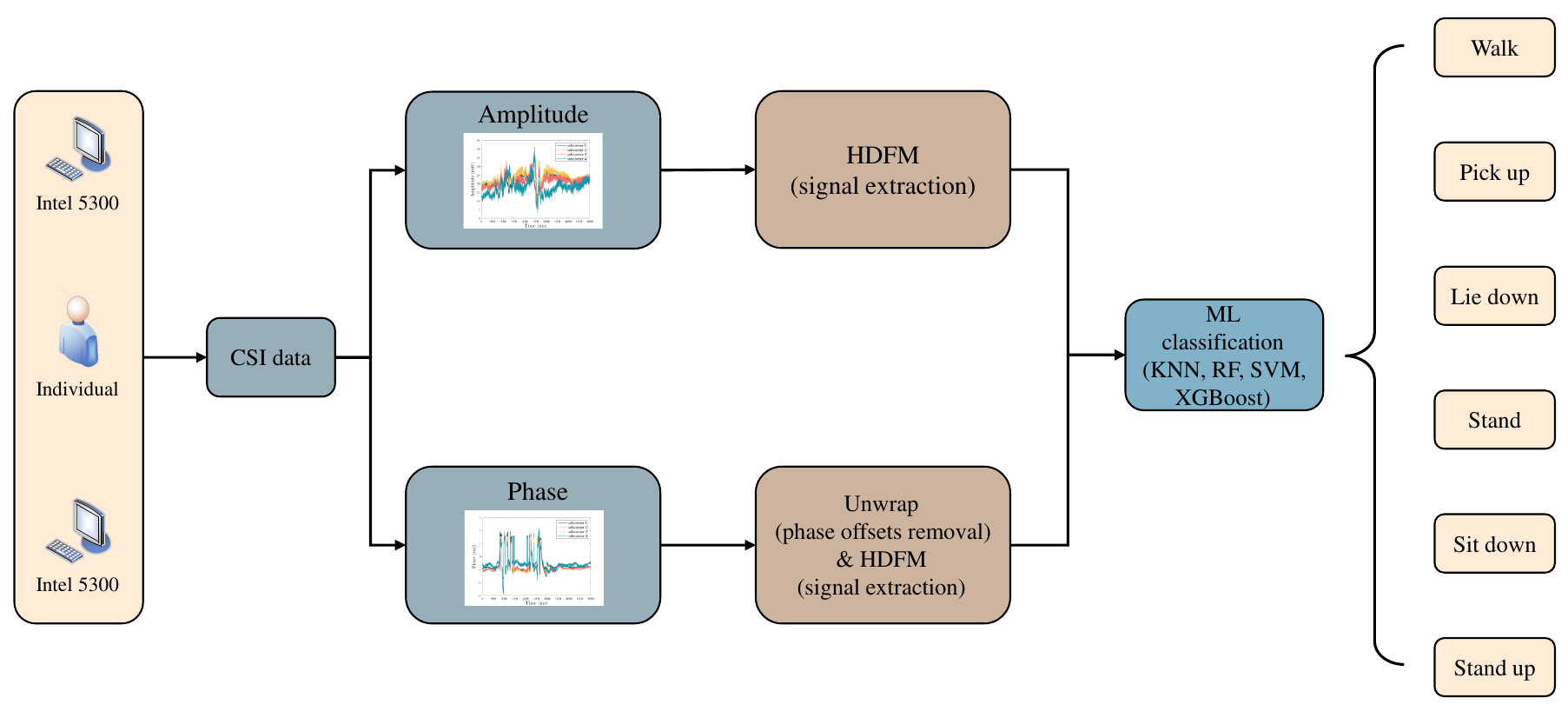}
    \caption{Overall architecture of the proposed HAR system.}
    \label{F3}
\end{figure*}

The basic steps involved in the proposed HAR model, are shown in Fig.~\ref{F3} and are described below.
\begin{itemize}
\item[$\bullet$] \textbf{CSI acquisition:} We collected CSI data for six human activity classes: lying down, picking up, sitting down, standing, standing up, and walking. This data was obtained using two computers equipped with Intel 5300 NICs, capable of capturing CSI data from 30 OFDM subcarriers at a 5 GHz center frequency and 20 MHz bandwidth.

\item[$\bullet$] \textbf{CSI preprocessing:} CSI data, represented as a complex matrix, is transformed into amplitude and phase information. This work utilizes both CSI amplitude and phase as features. Due to environmental effects, CSI amplitude data are often noisy and redundant. To address this, we propose the High-Dimensional Factor Model (HDFM) technique for signal extraction. This yields feature vectors, capturing subcarrier behavior. Similarly, CSI phase information undergoes signal extraction using HDFM, resulting in feature vectors after achieving the calibrated phase.

\item[$\bullet$] \textbf{Model training:} We employed four different machine learning classification models to validate the effectiveness of the HDFM model in dimensionality reduction.
\end{itemize}

\subsection{High-Dimensional Factor Models}
Consider a factor model that is as follows. For $i=1,2,\cdots,N$ and $t=1,2,\cdots,T$,
\begin{equation}
\label{factormodel}
\mathbf{R}_{it}=\sum_{j=1}^{p}\mathbf{L}_{ij}\mathbf{F}_{jt}+\mathbf{U}_{it}
\end{equation}
where $\mathbf{R}_{it}$ represents the observed data of $i$-th element at time $t$, $\mathbf{F}_{jt}$ denotes the $j$-th factor at time $t$, and $\mathbf{L}_{ij}$ represents the loading of $j$-th factor on the $i$-th cross-sectional element. Additionally, $\mathbf{U}_{it}$ stands for additive random noise or the idiosyncratic part that cannot be explained by the temporal factor $\mathbf{F}_{jt}$.

In this study, we evaluate the effectiveness of the Marčenko-Pastur (M-P) law\cite{marchenko1967distribution} in explaining the eigenvalue distribution of residuals $\mathbf{U}_{it}$ resulting from factor removal. Our analysis focuses on synthetic data, which is generated using the following model, with a fixed dimension of $N=400$ and $T=1000$.
\begin{equation}
\label{factormodel}
\mathbf{R}_{it}^{syn}=\sum_{j=1}^{p}\mathbf{L}_{ij}^{syn}\mathbf{F}_{jt}^{syn}+\mathbf{U}_{it}^{syn}
\end{equation}
where $\mathbf{L}_{ij}^{syn}, \mathbf{U}_{it}^{syn}\sim \mathcal{N}(0,1)$ are independent, $\mathbf{F}_{jt}^{syn}$ is generated by $\sin(2\pi f_j t)$ and the true number of factors, denoted as $p$, is set to be $3$ in our analysis.

Next, for both $\mathbf{R}_{it}$ and $\mathbf{R}_{it}^{syn}$, $p$-level residuals are created by factor removal through principal components:
\begin{equation}
\hat{\mathbf{U}}^{(p)} = \mathbf{R} - \hat{\mathbf{L}}^{(p)} \hat{\mathbf{F}}^{(p)}
\end{equation}
where $\hat{\mathbf{L}}^{(p)} \hat{\mathbf{F}}^{(p)}$ represents the estimated common factor obtained from $p$ principal components. Our focus lies in examining the eigenvalue distribution of the covariance matrix of residuals $\hat{\mathbf{U}}^{(p)}$:
\begin{equation}
\hat{\mathbf{C}}^{(p)} = \frac{1}{T} \hat{\mathbf{U}}^{(p)} \hat{\mathbf{U}}^{(p) T}
\end{equation}

\begin{figure}[tbp]
\centering
\subfloat[no factors removed]{
\includegraphics[width=0.48\columnwidth]{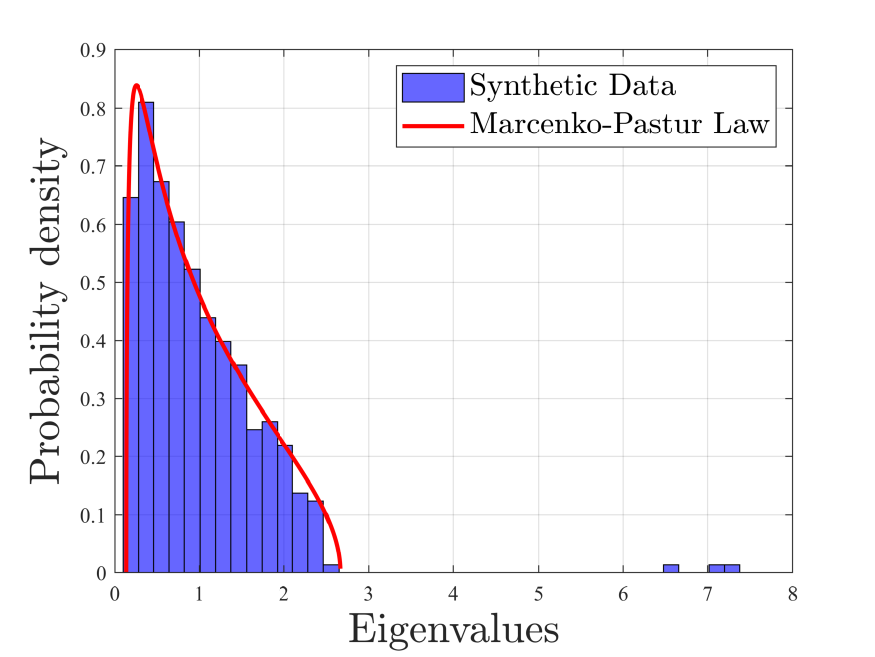}
}
\subfloat[one factor removed]{
\includegraphics[width=0.48\columnwidth]{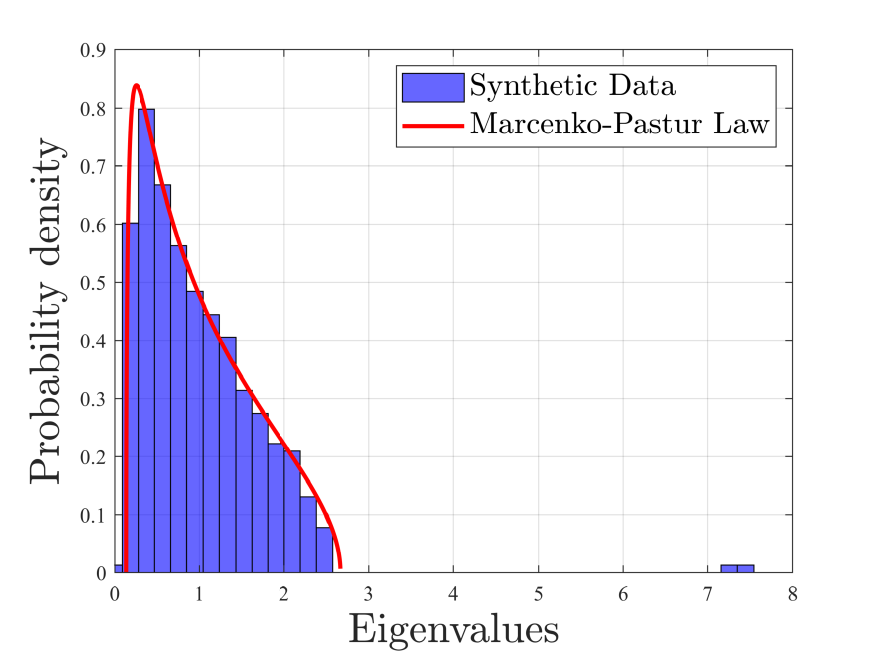}
}
\\
\subfloat[two factors removed]{
\includegraphics[width=0.48\columnwidth]{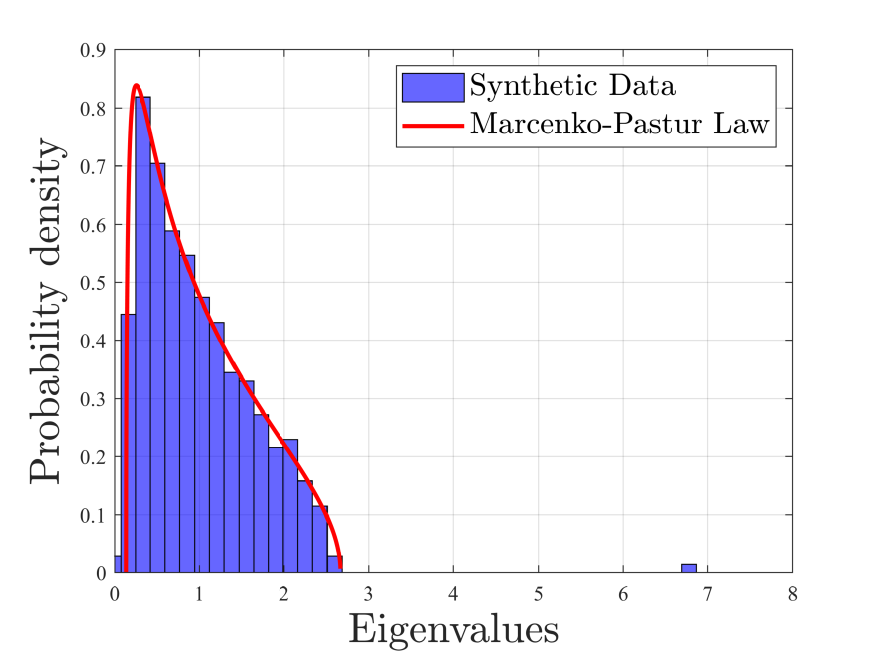}
}
\subfloat[three factors removed]{
\includegraphics[width=0.48\columnwidth]{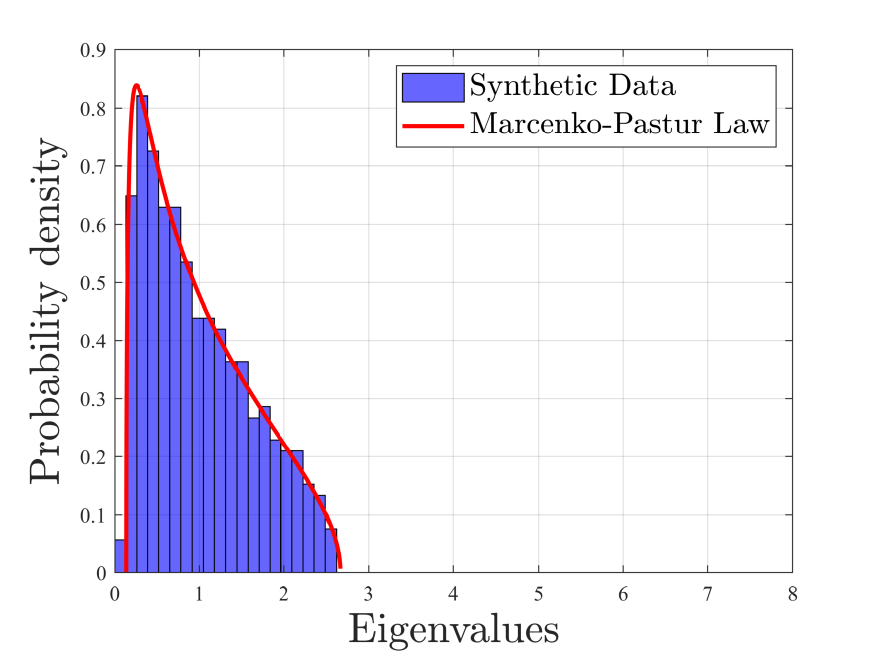}
}
\caption{The eigenvalue distribution of the covariance matrix of residual from synthetic data is studied with the removal of a few principal components. Specifically, removing the three factors leads to the disappearance of associated spikes in the eigenvalue distribution. The remaining bulk component closely conforms to the M-P law.}
\label{F4}
\end{figure}

Fig.~\ref{F4} illustrates the eigenvalue distribution of the residuals. The plot shows an empirical spectrum with a bulk component attributed to random noise and isolated spikes indicative of signals. In the synthetic data spectrum, three prominent spikes correspond to the three factors generated. Upon removing the correct number of factors, the residual spectral density converges to the M-P law.

\begin{figure*}[tbp]
    \centering
    \subfloat[PCA-KNN]{
		\includegraphics[width=.5\columnwidth]{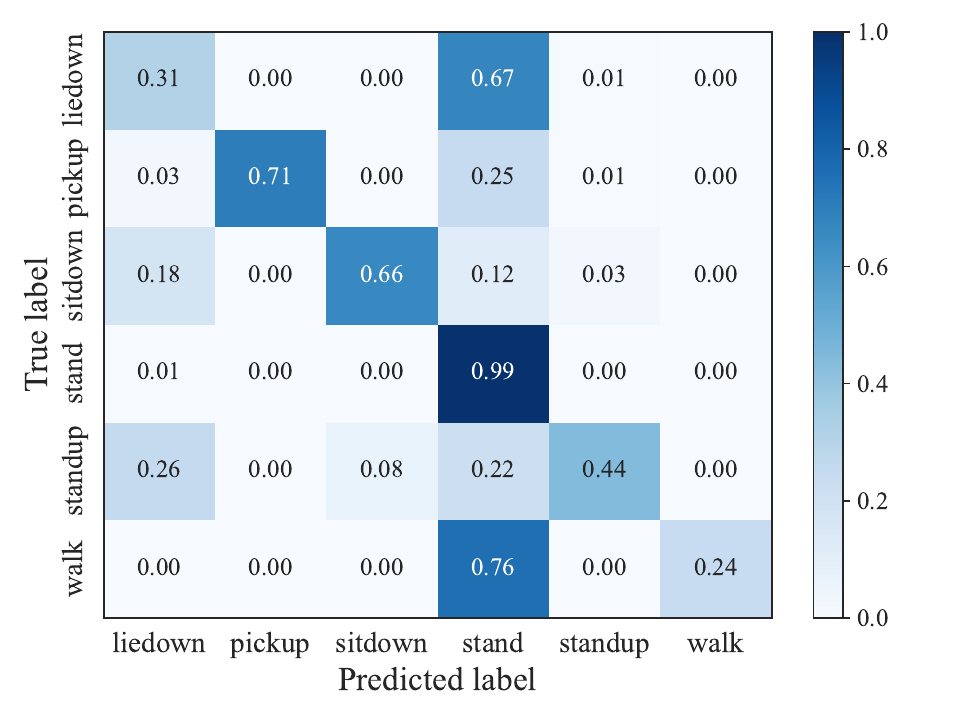}}
    \subfloat[PCA-RF]{
		\includegraphics[width=.5\columnwidth]{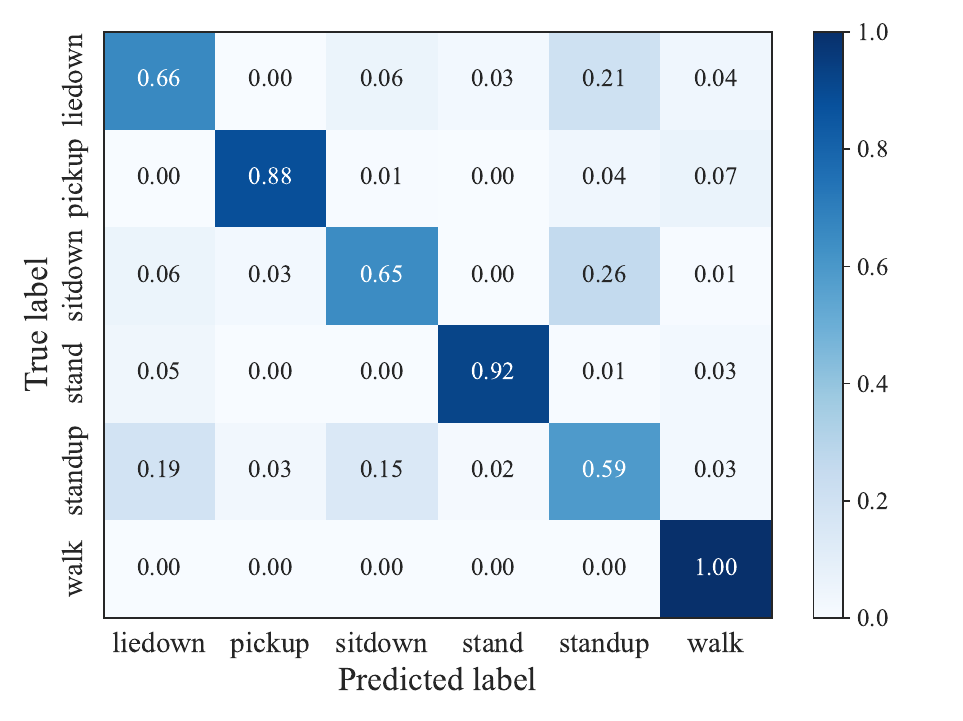}}
    \subfloat[PCA-XGBoost]{
		\includegraphics[width=.5\columnwidth]{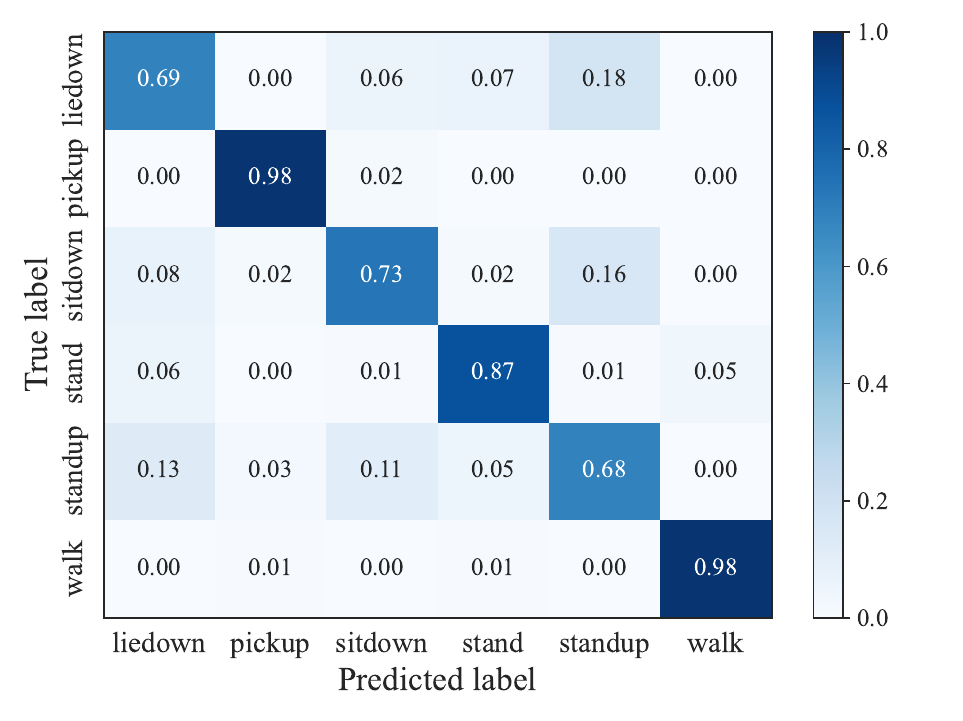}}
    \subfloat[PCA-SVM]{
		\includegraphics[width=.5\columnwidth]{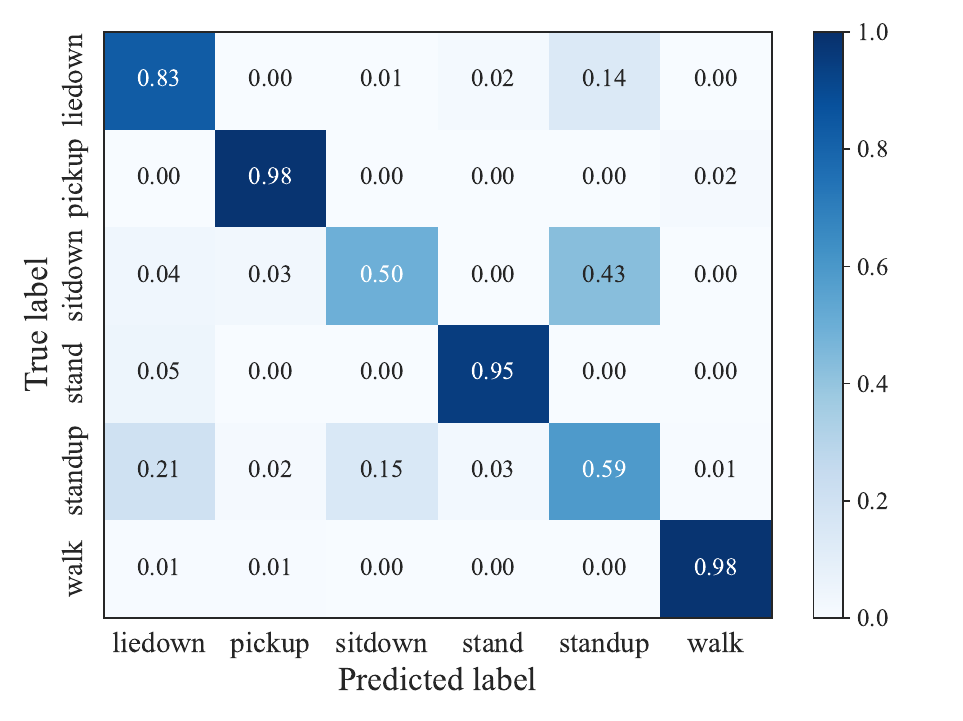}}
    \\
    \subfloat[HDFM-KNN]{
		\includegraphics[width=.5\columnwidth]{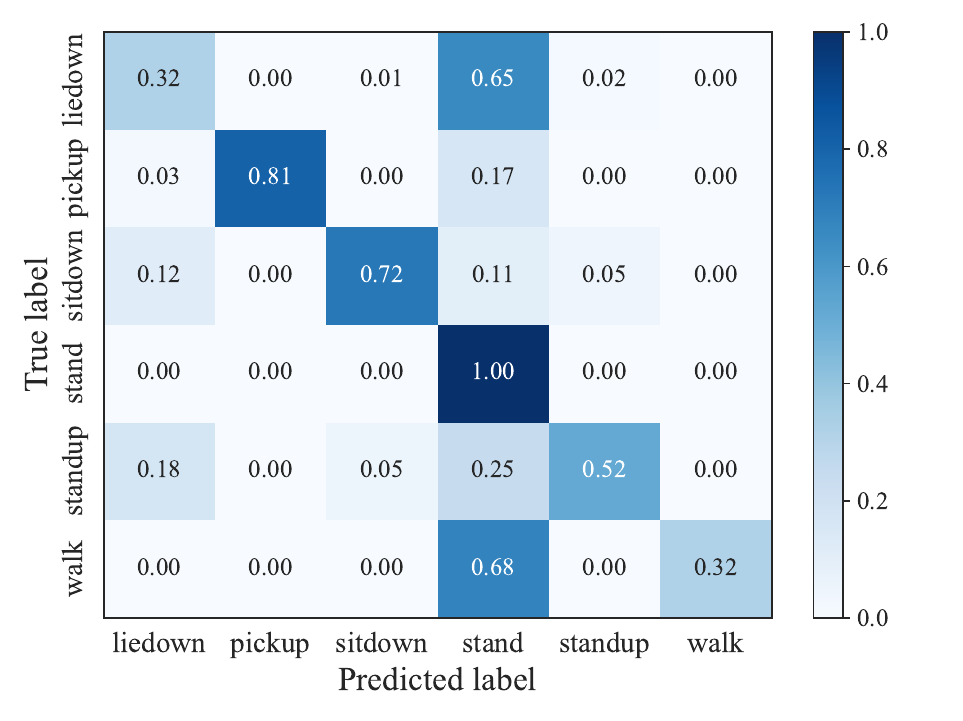}}
    \subfloat[HDFM-RF]{
		\includegraphics[width=.5\columnwidth]{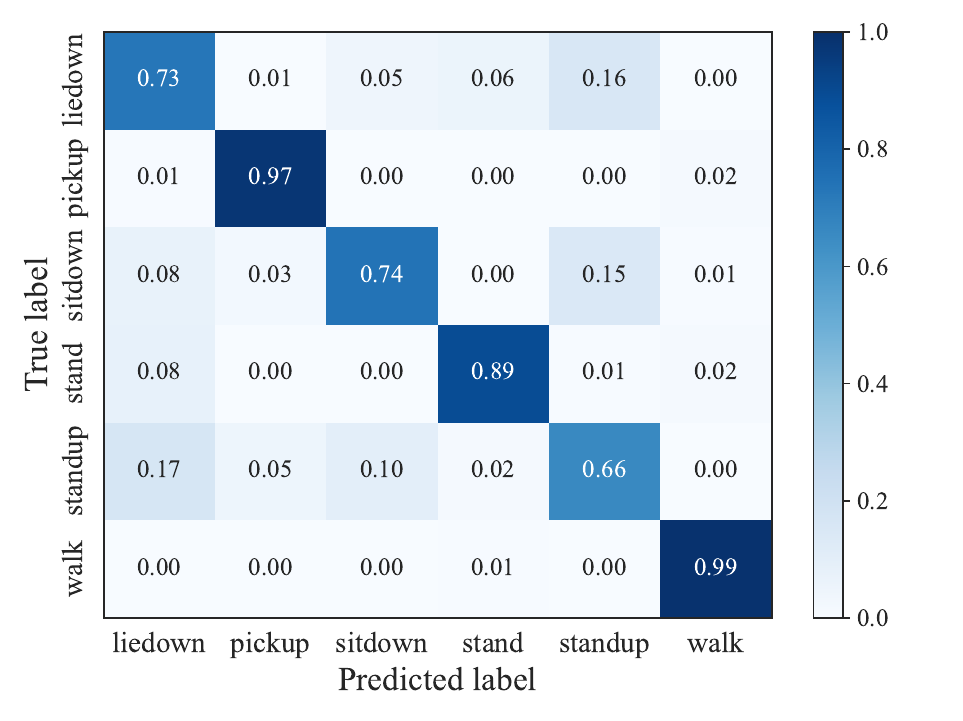}}
    \subfloat[HDFM-XGBoost]{
		\includegraphics[width=.5\columnwidth]{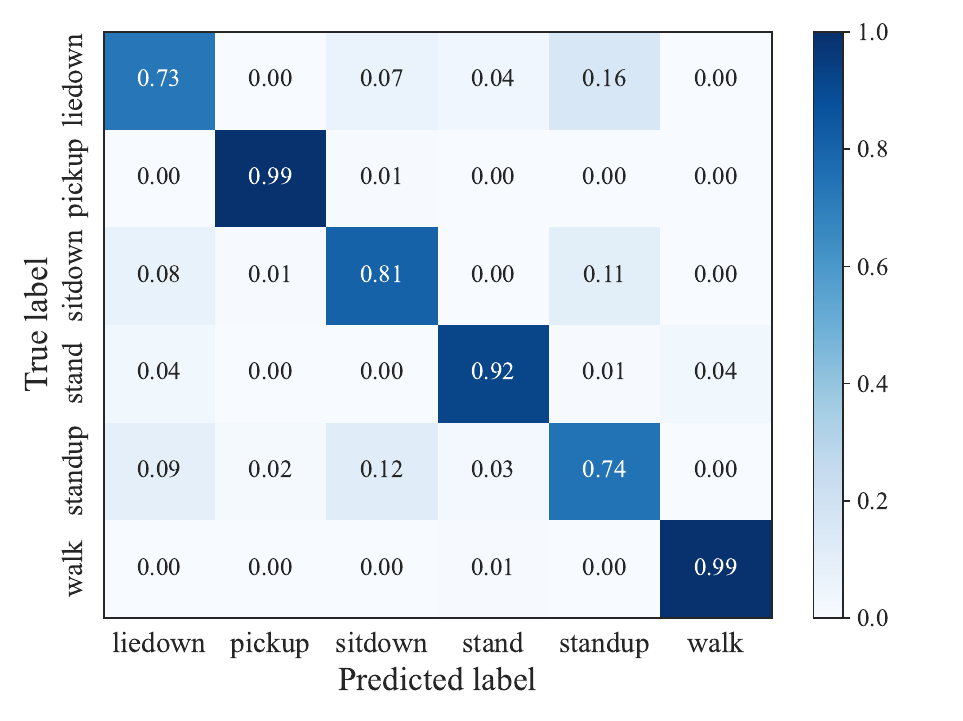}}
    \subfloat[HDFM-SVM]{
		\includegraphics[width=.5\columnwidth]{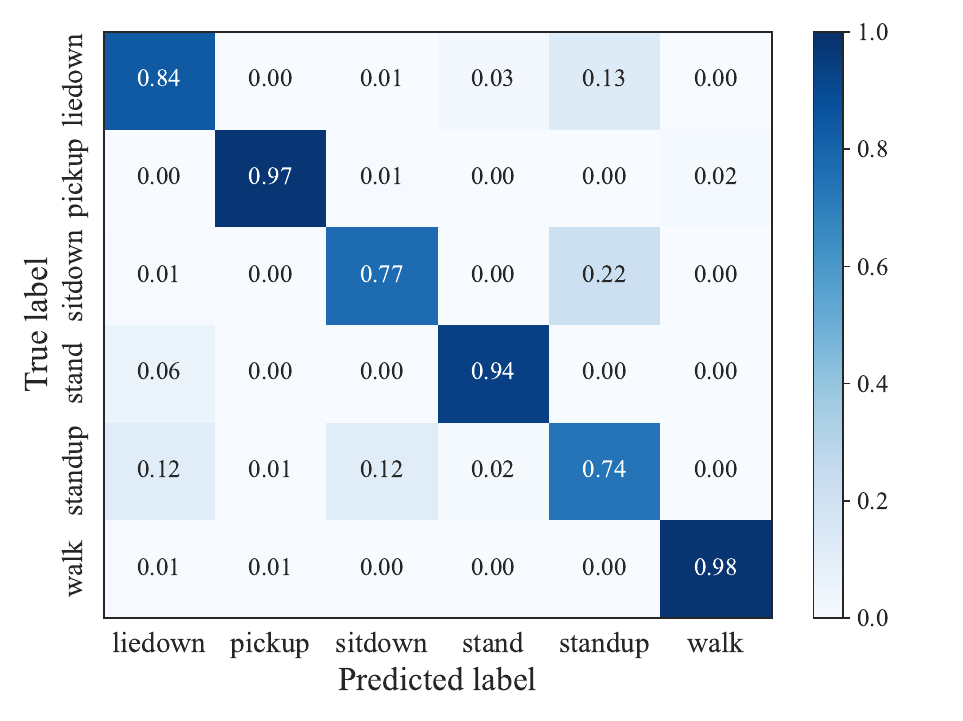}}
    \caption{Comparison of confusion matrices of different machine learning classification methods. Top line: Features extracted by PCA. Bottom line: Features extracted by HDFM.}
    \label{F5}
\end{figure*}

We employ the proposed method on CSI amplitude data to determine the number of factors based on the spectral distribution of $\mathbf{C}_{real}^{(p)}$. In our algorithm, we iteratively subtract factors by adjusting the value of $p$ until the bulk spectrum of the residuals using real data closely approximates the M-P law. The $p$ largest principal components extracted from real data are utilized as features for classification.

\section{Experimental setup and results}
\subsection{Experimental setup}
In this experiment, we utilize Intel 5300 NICs running on an Ubuntu 16.04 system. This setup allows for the reception of CSI packets using the CSI tool. The experiment involves a transmitter-receiver pair equipped with Intel 5300 NICs. Amplitude and phase variations in the CSI data are considered for the classification of six types of movements: lying down, picking up, sitting down, standing, standing up, and walking. Each action has a duration of 5 seconds and is sampled at a rate of 1 KHz, resulting in a total of 5000 sample points. Six different individuals perform each of the six actions, and each action is repeated 100 times. This yields a total of 600 sets of sample data.

Upon completion of the CSI data collection, we proceed to process the CSI data into magnitude and phase components. For the CSI magnitude data, we utilize the proposed algorithm, adjusting the value of $p$. Then we employ $p$ temporal factors as classification features. Regarding the CSI phase data, we initially apply the preprocessing algorithm outlined in Sec.~\ref{section2} to transform the raw phase data into the calibrated phase values. Similarly, we employ HDFM for signal extraction.

The features extracted from CSI data are utilized as input for machine learning models to classify human activities. In this study, we employ four distinct machine learning models for activity classification, namely, K-nearest Neighbor (KNN), Random Forest (RF), Extreme Gradient Boosting (XGBoost), and Support Vector Machine (SVM).

\subsection{Experimental results}
We apply both the PCA and HDFM methods to the datasets collected from our system, extracting distinct features as inputs for the classification algorithm. The performance is assessed using several parameters: Accuracy, Precision, Recall, and F1-score. To evaluate the performance of the proposed HDFM method, we compare and train several standard machine learning algorithms, including KNN, RF, XGBoost, and SVM. The trained confusion matrix is presented in Fig.\ref{F5}, highlighting that HDFM-based methods achieve higher accuracy in classifying the six actions compared to PCA-based approaches. Additionally, we evaluate the overall performance of various algorithms, as detailed in Table\ref{table1}, including metrics such as Recall, Precision, and F1-score. Across HDFM-based classification algorithms, there is an accuracy improvement ranging from 4.03\% to 6.81\% when compared to PCA-based methods.

\begin{table}[htbp]
\centering
\caption{The recognition performances of all the methods.}
\begin{tabular}{clcccc}
\toprule
\multicolumn{2}{c}{Method} & Accuracy & Recall & Precision & F1-score \\
\midrule
\multirow{2}{*}{KNN}       & PCA  & 0.5514   & 0.5602 & 0.7552    & 0.5680   \\
                           & HDFM & 0.6042   & 0.6147 & 0.7789    & 0.6224   \\
\multirow{2}{*}{RF}        & PCA  & 0.7736   & 0.7830 & 0.7837    & 0.7805   \\
                           & HDFM & 0.8236   & 0.8305 & 0.8283    & 0.8288   \\
\multirow{2}{*}{XGBoost}   & PCA  & 0.8153   & 0.8234 & 0.8190    & 0.8204   \\
                           & HDFM & 0.8556   & 0.8632 & 0.8593    & 0.8606   \\
\multirow{2}{*}{SVM}       & PCA  & 0.8014   & 0.8043 & 0.8133    & 0.8033   \\
                           & HDFM & 0.8694   & 0.8730 & 0.8768    & 0.8742   \\
\bottomrule
\end{tabular}
\label{table1}
\end{table}

In summary, it is evident from the table that Wi-Fi sensing when utilizing the proposed HDFM method, achieves superior recognition accuracy compared to other benchmark methods.

\section{Conclusion}
In this paper, we propose a signal extraction method based on Wi-Fi CSI. It exhibits superior performance when dealing with CSI data. Then the experimental results show that the proposed method has higher accuracy compared with the traditional PCA method.

\section*{Appendix}
Let $\mathbf{X}=\{ x_{ij} \}$ be a $N \times T$ random matrix, whose entires are independent identically distributed (i.i.d.) variables with the mean $\mu(x)=0$ and the variance $\sigma^2(x) < \infty $. The corresponding covariance matrix is defined as $\mathbf{C} = \frac{1}{T} \mathbf{X}\mathbf{X}^H$. As $N, T \longrightarrow \infty $ but $c = \frac{N}{T} \in (0, 1]$, according to M-P law, the empirical spectrum distribution of $\mathbf{C}$ converges to the limit with probability density function (PDF)
\begin{equation}
f(x)=\begin{cases}
\frac{1}{2\pi c \sigma^2 x} \sqrt{(b-x)(x-1)}, & a \leq x \leq b \\
0, & \mathrm{others}
\end{cases}
\end{equation}
where $a = \sigma^2(1-\sqrt{c})^2$, $b=\sigma^2(1+\sqrt{c})^2$.

\section*{Acknowledgment}
This work was supported by the Nation Natural Science Foundation of China under Grant No.12141107.


\begin{thebibliography}{10}

\bibitem{jovanov2019wearables}
E.~Jovanov, ``Wearables meet iot: Synergistic personal area networks (spans),'' {\em Sensors}, vol.~19, no.~19, p.~4295, 2019.

\bibitem{hassan2018robust}
M.~M. Hassan, M.~Z. Uddin, A.~Mohamed, and A.~Almogren, ``A robust human activity recognition system using smartphone sensors and deep learning,'' {\em Future Generation Computer Systems}, vol.~81, pp.~307--313, 2018.

\bibitem{zhou2020deep}
X.~Zhou, W.~Liang, I.~Kevin, K.~Wang, H.~Wang, L.~T. Yang, and Q.~Jin, ``Deep-learning-enhanced human activity recognition for internet of healthcare things,'' {\em IEEE Internet of Things Journal}, vol.~7, no.~7, pp.~6429--6438, 2020.

\bibitem{zhao2022human}
Y.~Zhao, H.~Zhou, S.~Lu, Y.~Liu, X.~An, and Q.~Liu, ``Human activity recognition based on non-contact radar data and improved pca method,'' {\em Applied Sciences}, vol.~12, no.~14, p.~7124, 2022.

\bibitem{lara2012survey}
O.~D. Lara and M.~A. Labrador, ``A survey on human activity recognition using wearable sensors,'' {\em IEEE communications surveys \& tutorials}, vol.~15, no.~3, pp.~1192--1209, 2012.

\bibitem{xu2013exploring}
X.~Xu, J.~Tang, X.~Zhang, X.~Liu, H.~Zhang, and Y.~Qiu, ``Exploring techniques for vision based human activity recognition: Methods, systems, and evaluation,'' {\em sensors}, vol.~13, no.~2, pp.~1635--1650, 2013.

\bibitem{kim2019vision}
K.~Kim, A.~Jalal, and M.~Mahmood, ``Vision-based human activity recognition system using depth silhouettes: A smart home system for monitoring the residents,'' {\em Journal of Electrical Engineering \& Technology}, vol.~14, pp.~2567--2573, 2019.

\bibitem{bhat2018online}
G.~Bhat, R.~Deb, V.~V. Chaurasia, H.~Shill, and U.~Y. Ogras, ``Online human activity recognition using low-power wearable devices,'' in {\em 2018 IEEE/ACM International Conference on Computer-Aided Design (ICCAD)}, pp.~1--8, IEEE, 2018.

\bibitem{qiu2022multi}
S.~Qiu, H.~Zhao, N.~Jiang, Z.~Wang, L.~Liu, Y.~An, H.~Zhao, X.~Miao, R.~Liu, and G.~Fortino, ``Multi-sensor information fusion based on machine learning for real applications in human activity recognition: State-of-the-art and research challenges,'' {\em Information Fusion}, vol.~80, pp.~241--265, 2022.

\bibitem{zhu2022continuous}
S.~Zhu, R.~G. Guendel, A.~Yarovoy, and F.~Fioranelli, ``Continuous human activity recognition with distributed radar sensor networks and cnn--rnn architectures,'' {\em IEEE Transactions on Geoscience and Remote Sensing}, vol.~60, pp.~1--15, 2022.

\bibitem{oguntala2019smartwall}
G.~A. Oguntala, R.~A. Abd-Alhameed, N.~T. Ali, Y.-F. Hu, J.~M. Noras, N.~N. Eya, I.~Elfergani, and J.~Rodriguez, ``Smartwall: Novel rfid-enabled ambient human activity recognition using machine learning for unobtrusive health monitoring,'' {\em IEEE Access}, vol.~7, pp.~68022--68033, 2019.

\bibitem{wang2017device}
W.~Wang, A.~X. Liu, M.~Shahzad, K.~Ling, and S.~Lu, ``Device-free human activity recognition using commercial wifi devices,'' {\em IEEE Journal on Selected Areas in Communications}, vol.~35, no.~5, pp.~1118--1131, 2017.

\bibitem{yan2019wiact}
H.~Yan, Y.~Zhang, Y.~Wang, and K.~Xu, ``Wiact: A passive wifi-based human activity recognition system,'' {\em IEEE Sensors Journal}, vol.~20, no.~1, pp.~296--305, 2019.

\bibitem{islam2022human}
M.~M. Islam, S.~Nooruddin, F.~Karray, and G.~Muhammad, ``Human activity recognition using tools of convolutional neural networks: A state of the art review, data sets, challenges, and future prospects,'' {\em Computers in Biology and Medicine}, p.~106060, 2022.

\bibitem{yang2013rssi}
Z.~Yang, Z.~Zhou, and Y.~Liu, ``From rssi to csi: Indoor localization via channel response,'' {\em ACM Computing Surveys (CSUR)}, vol.~46, no.~2, pp.~1--32, 2013.

\bibitem{halperin2011tool}
D.~Halperin, W.~Hu, A.~Sheth, and D.~Wetherall, ``Tool release: Gathering 802.11 n traces with channel state information,'' {\em ACM SIGCOMM computer communication review}, vol.~41, no.~1, pp.~53--53, 2011.

\bibitem{sun2022human}
Z.~Sun, Q.~Ke, H.~Rahmani, M.~Bennamoun, G.~Wang, and J.~Liu, ``Human action recognition from various data modalities: A review,'' {\em IEEE transactions on pattern analysis and machine intelligence}, 2022.

\bibitem{sharma2022review}
V.~Sharma, M.~Gupta, A.~K. Pandey, D.~Mishra, and A.~Kumar, ``A review of deep learning-based human activity recognition on benchmark video datasets,'' {\em Applied Artificial Intelligence}, vol.~36, no.~1, p.~2093705, 2022.

\bibitem{zhang2022deep}
S.~Zhang, Y.~Li, S.~Zhang, F.~Shahabi, S.~Xia, Y.~Deng, and N.~Alshurafa, ``Deep learning in human activity recognition with wearable sensors: A review on advances,'' {\em Sensors}, vol.~22, no.~4, p.~1476, 2022.

\bibitem{he2023robust}
Z.~He, X.~Zhang, Y.~Wang, Y.~Lin, G.~Gui, and H.~Gacanin, ``A robust csi-based wi-fi passive sensing method using attention mechanism deep learning,'' {\em IEEE Internet of Things Journal}, 2023.

\bibitem{chen2023cross}
C.~Chen, G.~Zhou, and Y.~Lin, ``Cross-domain wifi sensing with channel state information: A survey,'' {\em ACM Computing Surveys}, vol.~55, no.~11, pp.~1--37, 2023.

\bibitem{ma2019wifi}
Y.~Ma, G.~Zhou, and S.~Wang, ``Wifi sensing with channel state information: A survey,'' {\em ACM Computing Surveys (CSUR)}, vol.~52, no.~3, pp.~1--36, 2019.

\bibitem{wang2015phasefi}
X.~Wang, L.~Gao, and S.~Mao, ``Phasefi: Phase fingerprinting for indoor localization with a deep learning approach,'' in {\em 2015 IEEE Global Communications Conference (GLOBECOM)}, pp.~1--6, IEEE, 2015.

\bibitem{lu2019towards}
Y.~Lu, S.~Lv, and X.~Wang, ``Towards location independent gesture recognition with commodity wifi devices,'' {\em Electronics}, vol.~8, no.~10, p.~1069, 2019.

\bibitem{wang2015understanding}
W.~Wang, A.~X. Liu, M.~Shahzad, K.~Ling, and S.~Lu, ``Understanding and modeling of wifi signal based human activity recognition,'' in {\em Proceedings of the 21st annual international conference on mobile computing and networking}, pp.~65--76, 2015.

\bibitem{zhang2019widigr}
L.~Zhang, C.~Wang, M.~Ma, and D.~Zhang, ``Widigr: Direction-independent gait recognition system using commercial wi-fi devices,'' {\em IEEE Internet of Things Journal}, vol.~7, no.~2, pp.~1178--1191, 2019.

\bibitem{tan2020enabling}
S.~Tan, J.~Yang, and Y.~Chen, ``Enabling fine-grained finger gesture recognition on commodity wifi devices,'' {\em IEEE Transactions on Mobile Computing}, vol.~21, no.~8, pp.~2789--2802, 2020.

\bibitem{ma2019practical}
X.~Ma, Y.~Zhao, L.~Zhang, Q.~Gao, M.~Pan, and J.~Wang, ``Practical device-free gesture recognition using wifi signals based on metalearning,'' {\em IEEE Transactions on Industrial Informatics}, vol.~16, no.~1, pp.~228--237, 2019.

\bibitem{sheng2020deep}
B.~Sheng, F.~Xiao, L.~Sha, and L.~Sun, ``Deep spatial--temporal model based cross-scene action recognition using commodity wifi,'' {\em IEEE Internet of Things Journal}, vol.~7, no.~4, pp.~3592--3601, 2020.

\bibitem{perahia2013next}
E.~Perahia and R.~Stacey, {\em Next generation wireless LANs: 802.11 n and 802.11 ac}.
\newblock Cambridge university press, 2013.

\bibitem{qian2014pads}
K.~Qian, C.~Wu, Z.~Yang, Y.~Liu, and Z.~Zhou, ``Pads: Passive detection of moving targets with dynamic speed using phy layer information,'' in {\em 2014 20th IEEE international conference on parallel and distributed systems (ICPADS)}, pp.~1--8, IEEE, 2014.

\bibitem{marchenko1967distribution}
V.~A. Marchenko and L.~A. Pastur, ``Distribution of eigenvalues for some sets of random matrices,'' {\em Matematicheskii Sbornik}, vol.~114, no.~4, pp.~507--536, 1967.

\end{thebibliography}
\end{document}